\documentclass[aps,preprint,nofootinbib,floatfix]{revtex4-1}
\pdfoutput=1
\usepackage{graphicx} 
\usepackage{color}
\usepackage[latin1]{inputenc}
\usepackage{amsmath,amssymb}
\usepackage{hyperref}
\usepackage{epstopdf}
\usepackage{slashed}
\usepackage{geometry}                
\def\be{\begin{equation}}
\def\ee{\end{equation}}
\def\bea{\begin{eqnarray}}
\def\eea{\end{eqnarray}}
\newcommand{\gev}{~{\rm GeV}}

\begin{document}

\title{Muon-to-Electron Conversion in Mirror Fermion Model \\
with Electroweak Scale Non-Sterile Right-handed Neutrinos}

\author{P. Q. Hung$^{1,4}$, Trinh Le$^1$, Van Que Tran$^2$ and Tzu-Chiang Yuan$^{3,5}$
}

\affiliation{
$^1$Department of Physics, University of Virginia, Charlottesville, VA 22904-4714, USA\\
$^2$Department of Physics, National Taiwan Normal University, Taipei 116, Taiwan\\
$^3$Institute of Physics, Academia Sinica, Nangang, Taipei 11529, Taiwan \\
$^4$Center for Theoretical and Computational Physics, Hue University College of Education, Hue, Vietnam\\
$^5$Physics Division, National Center for Theoretical Sciences, Hsinchu, Taiwan
}

\date{\today}                                           

\begin{abstract}

The muon-to-electron conversion in nuclei like aluminum, titanium and gold is studied in the context of
a class of mirror fermion model with non-sterile right-handed neutrinos having mass at the electroweak scale.
At the limit of zero momentum transfer and large mirror lepton masses, 
we derive a simple formula to relate the conversion rate with the
on-shell radiative decay rate of muon into electron.
Current experimental limits (SINDRUM II) and projected sensitivities (Mu2e, 
COMET and PRISM) 
for the muon-to-electron conversion rates in various nuclei and latest limit from MEG for 
the radiative decay rate of muon into electron are used to put constraints on the parameter space of the model.  
Depending on the nuclei targets used in different experiments, 
for the mirror lepton mass in the range of 100 to 800 GeV, the sensitivities of the new Yukawa couplings
one can probe in the near future are in the range of one tenth to one hundred-thousandth,
depending on the mixing scenarios in the model.

\end{abstract}


\maketitle


\section{Introduction}\label{sec1}

As is well known, lepton flavor
is an accidental conserved quantity in Standard Model (SM) with strictly massless neutrinos.
For example, a muon never decays radiatively into an electron plus a photon and neutrinos 
do not oscillate in SM. However various experiments 
have now established firmly that neutrinos do oscillate from one flavor to another.
The common wisdom, motivated by the physics of $K-\overline K$ oscillation in the kaon system, is to give tiny masses with small mass differences to the various light neutrino species. 
Radiative decay of the muon into electron 
is then possible but with an unobservable rate highly suppressed 
with the minuscule neutrino masses~\cite{Petcov:1976ff,Ma:1980gm}. 
Searches for lepton flavor violating rare processes in high intensity experiments 
are thus important for new physics beyond the SM.

The most updated limit on $B(\mu \rightarrow e \gamma)$ is from MEG experiment~\cite{TheMEG:2016wtm} 
\be
\label{megcurrent}
      B(\mu \to e \gamma)  \le  4.2 \times 10^{-13} \; (90\% \, {\rm C.L.}) \,\hspace{10pt} {\rm(MEG \, 2016)} \, ,
\ee
and its projected improvement~\cite{Renga:2014xra} is 
\be
\label{megprojected}
      B(\mu \to e \gamma)  \sim  4 \times 10^{-14} \, .
\ee
Recent data from T2K experiment~\cite{IwamotoICHEPTalk} 
agrees well with the global analysis of neutrino oscillation data~\cite{Gonzalez-Garcia:2015qrr,Adamson:2016tbq,Adamson:2016xxw}, 
suggesting that the normal neutrino mass hierarchy (NH) with a CP violating Dirac phase $\delta_{\rm CP} \sim 3 \pi/2$ is slightly preferred. 
The best fit result for the central values of the PMNS matrix elements in 
the normal neutrino mass hierarchy can be extracted from~\cite{Gonzalez-Garcia:2015qrr} 
\bea
U^{\rm NH}_{\rm PMNS} =
\left(
\begin{array}{ccc}
0.8251 & 0.5453 & 0.08679 + 0.1195 i \\
-0.4568 + 0.0670 i & 0.5854 + 0.04428 i & 0.6649 \\
0.3171 + 0.07377 i & -0.5963 + 0.04875 i & 0.7322
\end{array}
\right) \; .
\label{UPMNSbestfits}
\eea

For the $\mu-e$ conversion in nuclei, the present experimental upper limits on the branching ratios were obtained by SINDRUM II experiment~\cite{Dohmen:1993plB,Bertl:2006up} for the targets titanium and gold, 
\bea
\label{sindrum2Ti}
  B(\mu^- + {\rm Ti}  \to e^- + {\rm Ti} ) & < & 4.3 \times 10^{-12} \; (90 \% \, {\rm C.L.)} \, , \\
\label{sindrum2Au}
  B(\mu^- + {\rm Au} \to e^- + {\rm Au}) & < & 7 \times 10^{-13} \; (90\% \, {\rm C.L.}) \, .
\eea
Significant improvements are expected for $\mu-e$ conversion at future experiments like Mu2e at Fermilab in US
and COMET at J-PARC in Japan. Projected sensitivities of $\mu-e$ conversion are~\cite{Bartoszek:2014mya, COMETphase1, Kuno:2013mha, Knoepfel:2013ouy,  Barlow:2011zza}
\bea
\label{COMETmu2eprojected}
  B(\mu^- + {\rm Al}  \to e^- + {\rm Al}) & < & 3 \times 10^{-17} \,\hspace{10pt}{\rm(Mu2e, COMET)}\, ,\\
  \label{mu2eIIPRISMprojected}
  B(\mu^- + {\rm Ti}  \to e^- + {\rm Ti}) & < &  10^{-18} \,\hspace{10pt}{\rm(Mu2e \, II, PRISM)}\, .
\eea
A positive signal of any of the above processes (or any process with charged lepton flavor violation (CLFV)) 
at the current or projected sensitivities of various high intensity experiments would be a clear indication of new physics as well, 
just like neutrino oscillations.
Given the fact that no new physics has showed up yet at the high energy frontier of the 
Large Hadron Collider (LHC), it is not a surprise that many recent works have been focused 
on new physics implication of CLFV in the high intensity frontier.
For a review on this topics and its possible connection with the muon anomaly, 
see~\cite{Lindner:2016bgg} and references therein.

In a recent work~\cite{Hung:2015hra}, we updated a previous calculation~\cite{Hung:2007ez} for 
the radiative process $\mu \to e \gamma$ in the mirror fermion model with electroweak scale non-sterile right-handed neutrinos~\cite{,Hung:2006ap} to 
an extended version~\cite{Hung:2015nva} where a horizontal $A_4$ symmetry in the lepton sector was imposed.
In this work we extend this previous analysis ~\cite{Hung:2015hra} to the $\mu-e$ conversion in nuclei, in particular for aluminum, gold and titanium. 

This paper is organized as follows. 
In Sec.~\ref{sec2}, after presenting some highlights of the crucial features of the extended mirror fermion model,
the calculation of $\mu  - e$ conversion in the model is presented.
In Sec.~\ref{sec3}, we derive a simple relation between the $\mu - e$ conversion rate and
the radiative decay rate of $\mu \to e\gamma$ in the limit of 
zero momentum transfer and large mirror lepton masses.
Numerical results are shown in Sec.~\ref{sec4}.
We summarize In Sec.~\ref{sec5}.
In Appendix A, we briefly review the effective Lagrangian~\cite{Kuno:1999jp, Kitano:2002mt} for 
describing $\mu - e$ conversion; 
and in Appendix B, we collect some useful formulas used in Sec.~\ref{sec3}.


\section{Mirror Fermion Model Calculation}
\label{sec2}

In this section, we first provide some highlights for the original mirror fermion model~\cite{Hung:2006ap} 
and its $A_4$ extension \cite{Hung:2015nva}. Then we compute the effective coupling constants  
induced at one loop level in the extended model for the $\mu - e$ conversion.


\subsection{Brief Review of Mirror Fermion Model}
\label{sec2a}

\subsubsection{Motivation}

The motivation of introducing mirror fermions in~\cite{Hung:2006ap} 
was manifold. First of all, it is aesthetically satisfactory
to have parity restoration at a higher
energy scale while the maximal parity violating interaction (V$-$A interaction) 
in SM can be emerged from spontaneous symmetry breaking. This is one of the main reasons for
various left-right symmetric models in the literature~\cite{pati-salam,moha-pati,goran-moha,goran}. 
Secondly, it is important to study non-perturbative effects in SM by discretizing 
it on the lattice. However it is well known that putting chiral fermion on the lattice is plagued by fermion doubling - an unavoidable consequence of 
the no-go theorem proved by Nielsen and Ninomiya~\cite{Nielsen:1981hk}. 
Sophisticated techniques like using Wilson fermion, Ginsparg-Wilson fermion, staggered fermion, or domain wall fermion {\it etc.}, which by violating at least one of the assumptions in the no-go theorem to get rid of the unwanted species,
are often employed to handle this problem in practice. 
For new physics model builders, it is attractive to
add mirror fermions to the SM which makes the theory becomes 
vector-like at a higher scale and hence one can avoid the fermion doubling 
problem if formulating on the lattice.
Chiral gauge anomalies will then be cancelled automatically in this class of models.
The third motivation is the electroweak scale non-sterile right-handed
neutrinos introduced in~\cite{Hung:2006ap}. For 
each generation, the right-handed neutrino is introduced together with 
a right-handed heavy charged fermion partner to form a SM SU(2) doublet. 
Similarly a left-handed heavy mirror charged lepton will be introduced for each
right-handed SM charged lepton. 
Majorana masses can then be given to these right-handed neutrinos via the vacuum expectation value (VEV) of a Higgs triplet with hypercharge $Y=2$ 
with mass at the electroweak scale, rather than the
grand unification scale in the usual scheme. Tiny Dirac masses can also be given via small VEVs of Higgs singlets with $Y=0$. 
This is the electroweak scale see-saw mechanism in mirror fermion model which is testable at the LHC~\cite{Chakdar:2016adj,Chakdar:2015sra}.

The original model in~\cite{Hung:2006ap} has been shown to be consistent with electroweak precision test data~\cite{Hoang:2013jfa} as well as the 125 GeV Higgs data from the LHC with an additional mirror Higgs doublet~\cite{Hoang:2014pda}. 
In~\cite{Hung:2015nva}, the original model was extended with a horizontal $A_4$ symmetry imposed in the lepton sector to address various issues of lepton mixings.
We briefly review this $A_4$ extension in the next subsection.

\subsubsection{Particle Content and Its $A_4$ Assignments}

The particle content of leptons and bosons of the model are shown in Table~\ref{particlecontent}.
The fields $l^M_{Ri}$ and $e^M_{Li}$ are the mirrors of the SM lepton doublet $l_{Li}$ and singlet $e_{Ri}$ respectively
for the $i$-th generation.
For the scalars, $\Phi_M$ is the mirror Higgs doublet of $\Phi$ introduced in~\cite{Hoang:2014pda}; $\xi$ and $\tilde \chi$ are the 
Georgi-Machacek triplets~\cite{Georgi:1985nv,Chanowitz:1985ug};
and $\phi_{0S}$ and $\vec \phi_S$ are singlets introduced 
in~\cite{Hung:2015nva}.
The $A_4$ assignments of these particles are listed at the second row in Table~\ref{particlecontent}.
Note that the scalar SU(2) singlet $\vec \phi_S$ is a $A_4$ triplet with its three components shown explicitly in the Table.

\begin{table}
\caption{The lepton and scalar sectors in the extended mirror model together with their assignments under the horizontal $A_4$ symmetry.}
\begin{tabular}{c|c|c|c|c||c|c|c|c|c|c}
\hline
&&&&&&&&&&\\
Fields & $ l_{Li}= \left( \begin{array}{c} \nu_L \\ e_L \end{array} \right)_i$
& $l^M_{Ri}= \left( \begin{array}{c} \nu_R \\ e^M_R \end{array} \right)_i$ & $e_{Ri}$ & $e^M_{Li}$ & $\phi_{0S}$ 
& $\vec\phi_{S} = \left( \begin{array}{c} \phi_{1S} \\ \phi_{2S} \\ \phi_{3S} \end{array} \right) $ & $\Phi$ & $\Phi_M$ & 
$\, \xi \, $ & $\, \tilde \chi \, $\\ 
&&&&&&&&&&\\
\hline
$SU(2)$ & \bf 2 & \bf 2 & \bf 1 & \bf 1 & \bf 1& \bf 1 & \bf 2 & \bf 2 & \bf 3 & \bf 3\\
\hline
$U(1)_Y$ & $-1/2$ & $-1/2$ & $-1$ & $-1$ & 0 & 0 & 1/2 & 1/2 & 0 & 2 \\
\hline
$A_4$ & \bf 3 & \bf 3 & \bf 3 & \bf 3 & \bf 1 & \bf 3 & \bf 1 & \bf 1 & \bf 1 & \bf 1\\
\hline
\end{tabular}
\label{particlecontent}
\end{table} 

The singlet scalars $\phi_{0S},\vec\phi_{S}$ are the only fields connecting the SM fermions and their mirror counterparts.
Recall that the tetrahedron symmetry group $A_4$ has four irreducible representations $\bf 1$, $\bf 1'$, $\bf 1''$, 
and $\bf 3$ with the following multiplication rule~\footnote{${\bf 3_1}$ 
is differ from ${\bf 3_2}$ because $A_4$ is nonabelian.}:
\begin{eqnarray}
{\bf 3} \times {\bf 3} & = &   {\bf 3_1} (23, 31, 12) + {\bf 3_2} (32, 13, 21) \nonumber \\
&+& {\bf 1} (11+ 22 + 33) + {\bf 1'} (11 + \omega^2 22 + \omega 33) 
+ {\bf 1''} (11 + \omega 22 + \omega^2 33)
\label{A4rules}
\end{eqnarray}
where $\omega = e^{2 \pi i/3}$.
In the gauge eigenbasis (fields with superscript 0), one can write down the following $A_4$ 
invariant Yukawa couplings,
\bea
- {\cal L}_S 
& = & g_{0S} \phi_{0S} (\overline{l^0_L}   l^{0M}_R)_{\bf 1} 
+ g_{1S} \vec \phi_S \cdot (\overline{l^0_L} \times l_R^{0M})_{\bf 3_1}
+ g_{2S} \vec \phi_S \cdot (\overline{l^0_L} \times l_R^{0M})_{\bf 3_2} 
+ {\rm H.c.} \nonumber \\
&+&  g^\prime_{0S} \phi_{0S} (\overline{e^0_R}   e^{0M}_L)_{\bf 1} 
+ g^\prime_{1S} \vec \phi_S \cdot (\overline{e^0_R} \times e_L^{0M})_{\bf 3_1}
+ g^\prime_{2S} \vec \phi_S \cdot (\overline{e^0_R} \times e_L^{0M})_{\bf 3_2} 
+ {\rm H.c.} \;\;
\label{A4Lang1}
\eea
As shown in \cite{Hung:2015nva}, after the scalar singlets develop VEVs with $v_0 =\langle \phi_{0S} \rangle$ 
and $v_i =\langle \phi_{iS} \rangle$, one obtains the neutrino mass matrix from the first line of (\ref{A4Lang1})
\be
M_\nu^{\rm Dirac} = 
\left(
  \begin{array}{cccc}
    g_{0S}v_0 & g_{1S}v_3 & g_{2S}v_2 \\
    g_{2S}v_3 & g_{0S}v_0 & g_{1S}v_1 \\
    g_{1S}v_2 & g_{2S}v_1 & g_{0S}v_0 \\
  \end{array}
\right) \, .
\ee
Hermiticity of the $M_\nu^{\rm Dirac}$ implies $g_{2S} = g^*_{1S}$.
Furthermore, if one assumes $v_i = v$, $M_\nu^{\rm Dirac}$ reduces to
\be
\label{mnu2}
M_\nu^{\rm Dirac} = 
\left(
  \begin{array}{cccc}
    g_{0S}v_0 & g_{1S}v & g^*_{1S}v \\
    g_{1S}^*v & g_{0S}v_0 & g_{1S}v \\
    g_{1S}v & g_{1S}^*v & g_{0S}v_0 \\
  \end{array}
\right) \, .
\ee
The above form of $M_\nu^{\rm Dirac}$ can be diagonalized by unitary transformation, 
{\it i.e.} $U^{\dagger}_\nu M_\nu^D U_\nu=M_\nu^{\rm Diag}$ 
with 
\be
\label{UCW}
U_\nu \equiv U_{\rm CW} =
\frac{1}{\sqrt{3}}
\left(
  \begin{array}{cccc}
  1 & 1 & 1 \\
  1 & \omega^2 & \omega \\
  1 & \omega & \omega^2 \\
  \end{array}
\right) \; ,
\ee
where $\omega$ is the same as in the multiplication rules of $A_4$ given in (\ref{A4rules}).
The matrix $U_{\rm CW}$ in (\ref{UCW}) was first discussed by Cabibbo~\cite{Cabibbo} and 
also by Wolfenstein~\cite{Wolfenstein} in the context
of CP violation in three generations of neutrino oscillations.
In recent years, advocating $A_4$ symmetry in the lepton sector 
was mainly due to Ma~\cite{ma}.

\vfill

\begin{table}
\caption{Matrix elements for the four auxiliary $M^k (k=0,1,2,3)$ where 
$\omega \equiv \exp(i 2\pi /3)$ and $g_{0S}$ and $g_{1S}$ are complex Yukawa couplings. 
$M^{\prime \, k}$ can be obtained from $M^{k}$ with the following substitutions
$g_{0S} \to g^\prime_{0S}$ and  $g_{1S} \to g^\prime_{1S}$.}
\begin{tabular}{|l|r|}
\hline
$M_{jn}^k$ & Value \\
\hline
\hline
$M^0_{12}, M^0_{13}, M^0_{21}, M^0_{23}, M^0_{31}, M^0_{32}$ & 0 \\
$M^0_{11},  M^0_{22}, M^0_{33}$ & $g_{0S}$ \\
$M^1_{11},  M^2_{11}, M^3_{11};\, M^1_{23}, M^1_{32}$  & $\frac{2}{3} \mathrm{Re} \left( g_{1S} \right)$ \\ 
$M^1_{22}, M^2_{22}, M^3_{22};\, M^1_{13},  M^1_{31}$  &  $\frac{2}{3} \mathrm{Re} \left( \omega^* g_{1S} \right)$  \\
$M^1_{33}, M^2_{33}, M^3_{33};\,M^1_{12}, M^1_{21}$  & $\frac{2}{3} \mathrm{Re} \left( \omega g_{1S} \right)$  \\ 
$M^2_{12}, M^3_{21}$ & $\frac{1}{3} \left( g_{1S} + \omega g^*_{1S} \right)$ \\
$M^3_{12}, M^2_{21}$ & $\frac{1}{3} \left( g^*_{1S} + \omega^* g_{1S} \right)$ \\ 
$M^2_{13},  M^3_{31}$ & $\frac{1}{3} \left( g_{1S} + \omega^* g^*_{1S} \right)$ \\
$M^3_{13},  M^2_{31}$ & $\frac{1}{3} \left( g^*_{1S} + \omega g_{1S} \right)$ \\
$M^2_{23}, M^3_{32}$ & $\frac{2 \omega^*}{3} \mathrm{Re} \left( g_{1S} \right)$ \\
$M^3_{23}, M^2_{32}$ & $\frac{2 \omega}{3} \mathrm{Re} \left( g_{1S} \right)$  \\
\hline
\end{tabular}
\label{M}
\end{table}

\subsubsection{Mixings}

Let $U^{l}_{L,R}$ and $U^{l^M}_{R,L}$ be the unitary matrices relating the gauge eigenstates 
and the mass eigenstates (fields without superscripts 0) defined as
\be
\label{eigenstate}
l^{0}_{L}= U^{l}_{L} l_{L} \; , \; e^{0}_{R}= U^{l}_{R} e_{R} \;, \;  
l^{M,0}_{R} =  U^{l^M}_{R} l^{M}_{R} \; , \; e^{M,0}_{L} =  U^{l^M}_{L} e^{M}_{L} \; .
\ee
Following~\cite{Hung:2015hra}, we express the Yukawa couplings in (\ref{A4Lang1}) 
as follows
\bea
\label{La}
{\mathcal L}^{l}_{S} & = & - \sum_{k=0}^3 \sum_{i,m=1}^3  \left( \bar{l}_{Li} \, {\cal U}^{L \, k}_{im} l^M_{R m}  
+ \bar{ e}_{R i}  \, {\cal U}^{R \, k}_{im} e^M_{Lm} \right) \phi_{kS} + {\rm H.c.} \; 
\eea
The coupling coefficients ${\cal U}^{L \, k}_{im}$ and ${\cal U}^{R \, k}_{im}$ are given by
\bea
\label{UL}
{\cal U}^{L \, k}_{im} 
& \equiv & \left ( U^\dagger_{\rm PMNS}  \cdot   M^k \cdot  U^{M}_{\rm PMNS} \right)_{im} \;\; , \nonumber \\
& = & \sum_{j,n = 1}^3 \left( U^\dagger_{\rm PMNS} \right)_{i j}   M^k_{jn}  
\left(  U^{M}_{\rm PMNS} \right)_{nm} \; \; , \\
{\cal U}^{R \, k}_{im} 
& \equiv & \left( U^{\prime \, \dagger}_{\rm PMNS} \cdot   M^{\prime \, k} \cdot U^{\prime \, M}_{\rm PMNS} \right)_{im} \; \; , \nonumber \\
& = & \sum_{j,n = 1}^3 \left( U^{\prime \, \dagger}_{\rm PMNS} \right)_{i j}   M^{\prime \, k}_{jn}  
\left(  U^{\prime \, M}_{\rm PMNS} \right)_{nm} \; \; ,
\label{UR}
\eea
where the matrix elements for the four auxiliary matrices $M^k (k=0,1,2,3)$ are listed in Table~\ref{M}, 
and $M^{\prime \, k}_{jn}$ can be obtained from $M^{k}_{jn}$ with the following substitutions
for the Yukawa couplings $g_{0S} \to g^\prime_{0S}$ and  $g_{1S} \to g^\prime_{1S}$;
$U_{\rm PMNS}$ is the usual neutrino mixing matrix defined as
\be
\label{PMNS}
U_{\rm PMNS}=  U_{\nu}^{\dagger} U^{l}_{L} \, ,
\ee
and its mirror and right-handed counter-parts  $U^{M}_{\rm PMNS} $, $U^\prime_{\rm PMNS}$ 
and $U^{\prime M}_{\rm PMNS} $  are defined analogously as
\be
\label{MPMNS}
U^{M}_{\rm PMNS} = U^{\dagger}_{\nu} U^{l^M}_R \, ,
\ee
\be
\label{UPMNSprime}
U^\prime_{\rm PMNS}=  U_{\nu}^{\dagger} U^{l}_{R} \, ,
\ee
and
\be
\label{UPMNSMirrorprime}
U^{\prime M}_{\rm PMNS} = U^{\dagger}_{\nu} U^{l^M}_L \,.
\ee


%
\begin{figure}[hbtp!]
\centering
\includegraphics[width=0.75\textwidth]{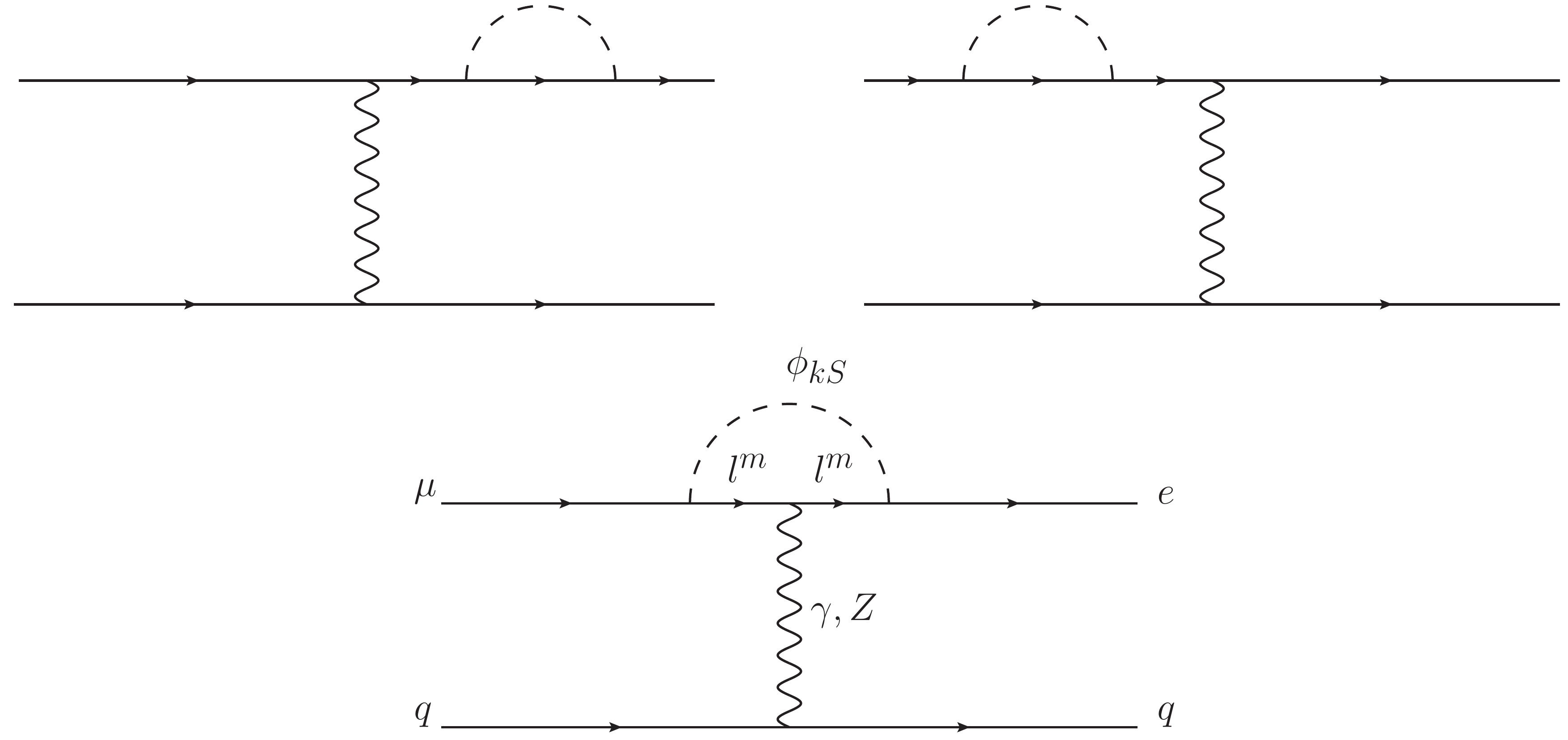} 
\caption{{\small One-loop induced Feynman diagrams from photon and $Z$ boson exchanges for $\mu - e$ conversion in electroweak-scale $\nu_R$ model}.}
\label{FeynDiags}
\end{figure}

\subsection{Photon Contributions and the Monopole and Dipole Form Factors}
\label{sec2b}

In this work we will focus on the contributions from 
the photon exchange as shown in the Feynman Diagrams of Fig.~\ref{FeynDiags}. 
We also compute the contributions from the $Z$-exchange but since they are suppressed by 
$m^2_\mu / m_Z^2$ we will not present them here.
The  invariant amplitude for $\mu^- (p) \to e^- (p') \gamma^*(q)$ with an off-shell photon 
can be parametrized as
\begin{equation}
\label{Mgamma}
i {\cal M}_{\gamma} = - e {\overline u}_e (p') i \Gamma^\mu_\gamma (q) u_{\mu}(p) A^*_\mu (q)
\end{equation} 
where $\Gamma^\mu_\gamma (q)$ has the following Lorentz and gauge invariant decomposition
\begin{equation}
\label{vertexgamma}
\Gamma^\mu_\gamma (q) =  
\left( f_{E0} (q^2) + \gamma_5 f_{M0} (q^2) \right) \left( \gamma^\mu - \frac{q_\mu \slash {\! \! \! q}}{q^2} \right)
+ \left( f_{M1} (q^2) + \gamma_5 f_{E1} (q^2) \right)  
\frac{i \sigma^{\mu\nu}q_\nu}{m_\mu} \; .
\end{equation}
The monopole form factors $f_{E0}$, $f_{M0}$ and the dipole form factors  
$f_{M1}$, $f_{E1}$ can be obtained by generalizing our previous on-shell 
calculation of $\mu \to e \gamma$ in the same model  \cite{Hung:2015hra} to the case of off-shell photon $\gamma^*$.
From the Feynman diagrams of Fig.~\ref{FeynDiags}, we obtain the following expressions 
\begin{eqnarray}
f_{E0,M0}(q^2) = & + &\frac{1}{32 \pi^2} \sum_{k,m}  \int_0^1 dx \int_0^{1-x} dy \left\{  \frac{x y q^2 }{\Delta_{km} (q^2)}
\left( {\cal U}^{L \, k}_{1m} \left( {\cal U}^{L \, k}_{2m} \right)^* \pm 
{\cal U}^{R \, k}_{1m}  \left( {\cal U}^{R\, k}_{2m} \right)^* \right) \right. \nonumber \\
&-& 
\left[ \log \left(\frac{\Delta_{km} (q^2)}{\Delta_{km} (0)} \right) - \left( M_m^2 \pm (1-x-y)^2 m_\mu m_e \right) 
\left( \Delta^{-1}_{km}(q^2) - \Delta^{-1}_{km}(0) \right)
\right] \nonumber \\
&& \;\;\; \times \left( {\cal U}^{L \, k}_{1m} \left( {\cal U}^{L \, k}_{2m} \right)^* \pm 
{\cal U}^{R \, k}_{1m}  \left( {\cal U}^{R\, k}_{2m} \right)^* \right) \nonumber \\
& + &   (1-x-y) (m_\mu \pm m_e ) M_m \left( \Delta^{-1}_{km}(q^2) - \Delta^{-1}_{km}(0) \right) \nonumber \\
&& \biggl. \;\;\;  \times \left( {\cal U}^{L \, k}_{1m} \left( {\cal U}^{R \, k}_{2m} \right)^* \pm 
{\cal U}^{R \, k}_{1m}  \left( {\cal U}^{L\, k}_{2m} \right)^* \right)
\biggr\}
\label{fEM0}
\end{eqnarray}
for the monopole form factors, and
\begin{eqnarray}
f_{M1,E1}(q^2) & = & - \frac{m_\mu}{32 \pi^2}
\sum_{k,m}  \int_0^1 dx \int_0^{1-x} dy \frac{1 }{\Delta_{km} (q^2)} \nonumber \\
&& \times \biggl\{  (1-x-y) \left( y m_\mu \pm x m_e \right) 
\left( {\cal U}^{L \, k}_{1m} \left( {\cal U}^{L \, k}_{2m} \right)^* \pm 
{\cal U}^{R \, k}_{1m}  \left( {\cal U}^{R\, k}_{2m} \right)^* \right) \biggr. \nonumber \\
&& \;\;\; \biggl. + (x+y) M_m 
\left( {\cal U}^{L \, k}_{1m} \left( {\cal U}^{R \, k}_{2m} \right)^* \pm 
{\cal U}^{R \, k}_{1m}  \left( {\cal U}^{L\, k}_{2m} \right)^* \right)
\biggr\}
\label{fEM1}
\end{eqnarray}
for the dipole form factors. Here, we have defined
\begin{equation}
\label{propagatorc}
\Delta_{km}(q^2) =  ( x+ y ) M_m^2 + ( 1 - x - y) ( m_k^2 - x m_e^2 - y m_\mu^2 ) - x y q^2 - i 0^+ \; ,
\end{equation}
where $m_k$ denotes the mass of scalar singlet $\phi_{kS}$ for $k=0,1,2, 3$
and $M_m$ the mass of mirror lepton $l^M_m$ for $m=1,2,3$.

At $q^2=0$, we have $f_{E0,M0}(0) = 0$ as one would expect. Thus the following reduced monopole 
form factors $\tilde f_{E0,M0}$ with an explicit factor
of $q^2$ extracted from  $f_{E0,M0}$ are often defined in the literature,
\begin{equation}
\label{reducedff}
f_{E0,M0} (q^2) = \frac{q^2}{m_\mu^2}\tilde f_{E0,M0} (q^2) \; .
\end{equation}
For small $q^2$, one can set 
${\tilde f}_{E0,M0}(q^2) \approx {\tilde f}_{E0,M0}(0)$ with
\begin{eqnarray}
{\tilde f}_{E0,M0}(0) & = &\frac{m_\mu^2}{32 \pi^2} \sum_{k,m}  \int_0^1 dx \int_0^{1-x} dy  \frac{x y }{\big(\Delta_{km} (0)\big)^2}\, \biggr\{ \left( {\cal U}^{L \, k}_{1m} \left( {\cal U}^{L \, k}_{2m} \right)^* \pm {\cal U}^{R \, k}_{1m}  \left( {\cal U}^{R\, k}_{2m} \right)^* \right)  \nonumber \\
 &&\hspace{80 pt} \times \Big( 2\Delta_{km} (0)   +  M_m^2 \pm (1-x-y)^2 m_\mu m_e \Big) \nonumber \\
& & \hspace{20 pt}+ \, \,\left( {\cal U}^{L \, k}_{1m} \left( {\cal U}^{R \, k}_{2m} \right)^* \pm {\cal U}^{R \, k}_{1m}  \left( {\cal U}^{L\, k}_{2m} \right)^* \right) (1-x-y) (m_\mu \pm m_e ) M_m\biggr\} . 
\label{reducedfEM0}
\end{eqnarray} 
The explicit factor of $q^2$ in (\ref{reducedff}) will cancel the $1/q^2$ of the photon propagator in Fig.~\ref{FeynDiags}.
This leads to four-fermion vector-vector interaction and hence  the reduced monopole form factors will
contribute to the effective coupling $C^{(q)}_{V(R,L)}$ in the effective Lagrangian of (\ref{Leff2}) in Appendix A.
We will discuss more about these four-fermion interactions in the next subsection.
 
At $q^2=0$, the contributions from the magnetic and electric dipole terms 
of (\ref{vertexgamma}) to the amplitude $\cal M_\gamma$ in (\ref{Mgamma}) can be
reproduced by the following effective Lagrangian
\begin{equation}
{\cal L}_{\gamma , \rm eff} = \frac{e}{2 m_\mu} \overline e \sigma^{\alpha\beta} 
\left( f_{M1}(0) + \gamma_5 f_{E1}(0) \right) \mu F_{\alpha\beta} + {\rm H.c.} \; .
\label{Leffgammadiople}
\end{equation}
Comparing (\ref{Leffgammadiople}) with the first line of the general form 
of the Lagrangian for $\mu - e$ conversion given in 
(\ref{Leff2}) in Appendix A, one can deduce
the dimensionless effective couplings $C_{DR,DL}$ as linear combinations of the 
static limit of the dipole form factors $f_{E1}$ and $f_{M1}$,
\begin{equation}
\frac{C_{DR,DL}}{\Lambda^2} =  \frac{e}{2 m_\mu^2} \left( \pm f_{E1} (0) - f_{M1}(0) \right) \; .
\label{CsubDRDL}
\end{equation}


\subsection{Four-Fermion Coupling Constants $C^{(q)}_{V(L,R)}$ - Photon Exchange}
\label{sec2c}

The amplitude for $\mu (p) q(k) \to e (p^\prime) q(k^\prime)$ 
from the monopole form factors of the photon exchange in Fig.~\ref{FeynDiags} can be obtained as
\begin{eqnarray}
\label{Ampgamma1}
{\cal M}_\gamma & = & - e^2 Q_q \overline{u_e}(p') \left(
f_{E0} (q^2) + f_{M0} (q^2) \gamma_5 \right) \left( \gamma_\mu - \frac{q_\mu \slash \!\!\! q}{q^2} \right) u_\mu (p)
\frac{1}{q^2} \overline{u_q}(k') \gamma^\mu u_q(k) \, , \, 
\end{eqnarray}
where $q=p-p'=k'-k$, and $f_{E0}$, $f_{M0}$ are given in (\ref{fEM0}).
The $q_\mu$ term in (\ref{Ampgamma1}) can be dropped due to quark current conservation.
As mentioned earlier, the $1/q^2$ of the photon propagator will be cancelled from a factor of $q^2$
in $f_{E0,M0}$. Thus in terms of the reduced form factors $\tilde f_{E0,M0}$ of (\ref{reducedff}), the amplitude 
${\cal M}_\gamma$ can be rewritten as
\begin{eqnarray}
\label{Ampgamma2}
{\cal M}_\gamma  = & - &\frac{e^2 Q_q}{m_\mu^2} \left[ \left( \tilde f_{E0} - \tilde f_{M0} \right) 
\overline{u_{Le}} (p') \gamma_\mu  u_{L\mu} (p) 
+
\left( \tilde f_{E0} + \tilde f_{M0} \right) 
\overline{u_{Re}} (p') \gamma_\mu  u_{R\mu} (p)
\right] \nonumber \\
& & \;\;\; \;\; \;\; \;  \times
\left[ \overline{ u_{Lq}}(k') \gamma^\mu u_{Lq}(k) + \overline{u_{Rq}}(k') \gamma^\mu u_{Rq}(k) \right] \;  ,
\end{eqnarray}
where $\tilde f_{E0,M0}$ are defined in (\ref{reducedfEM0}) for small $q^2$. 
At $q^2 = 0$, this amplitude can be reproduced by the following Fermi interaction
\begin{eqnarray}
\label{Leffgamma}
{\cal L}^\prime_{\gamma , \rm eff} & = & - \frac{e^2 Q_q}{m_\mu^2} 
\left[ 
\left( \tilde f_{E0}(0) - \tilde f_{M0}(0) \right) \overline{e_L} \gamma_\mu \mu_L + 
\left( \tilde f_{E0}(0) + \tilde f_{M0}(0) \right) \overline{e_R} \gamma_\mu \mu_R 
\right] \nonumber \\
&& \;\;\;\;\;\;\;\;\; \; \; \times \left[ \overline q \gamma^\mu q \right] \; .
\end{eqnarray}
By matching (\ref{Leffgamma}) with the second line of the general form 
of the Lagrangian for $\mu - e$ conversion given in (\ref{Leff2}) in Appendix A, we 
deduce the following relations for the dimensionless effective couplings 
$C^{(q)\gamma}_{V(L,R)}$
\begin{eqnarray}
\frac{C^{(q)\gamma}_{V(L,R)}}{\Lambda^2} & = & \frac{e^2 Q_q}{m_\mu^2} \left( \tilde f_{E0} (0) \mp \tilde f_{M0} (0) \right) \; . 
\label{CsubVLVR}
\end{eqnarray}
Note that we have the relation $C^{(u)\gamma}_{V(L,R)} = -2 C^{(d)\gamma}_{V(L,R)}$. This implies the vector effective couplings $\tilde C^{(n) \gamma }_{V(L,R)}$ for the neutron from the photon exchange are vanishing. This is expected 
since neutron carries no electric charge. 

We also note that for the photon contributions, 
only $C^{(q)}_{VR}$ and $C^{(q)}_{VL}$ are non-vanishing. 
Other four-fermion effective couplings will be non-vanishing only from $Z$ exchange, scalar exchange or box diagrams, which are negligible as compared with the photon exchange contributions.


\section{The relationship between $\mu-e$ Conversion and $\mu \to e \gamma$ }
\label{sec3}

Since the momentum transfer $q^2$ is expected to be quite small
in the $\mu-e$ conversion process in nuclei, we can make a Taylor expansion for the various form factors deduced in the previous section 
around $q^2 = 0$.
Thus for small $q^2$, we have
\begin{eqnarray}
\nonumber
\label{infEM}
f_{E0,M0}(q^2) & \approx & \frac{q^2}{32 \pi^2} \frac{1}{M_m^4}  \sum_{k,m}
  \biggl\{
\left( {\cal U}^{L \, k}_{1m} \left( {\cal U}^{L \, k}_{2m} \right)^* \pm 
{\cal U}^{R \, k}_{1m}  \left( {\cal U}^{R\, k}_{2m} \right)^* \right) \nonumber \\
&& \hspace{2.8cm} \times   \left[
M_m^2 \left({\cal I}(r_{km}) + 2 {\cal I}_{30}(r_{km}) \right)  \pm m_\mu m_e {\cal I}_{10}(r_{km}) \right] \\
&& \hspace{2.3cm}
+\left( {\cal U}^{L \, k}_{1m} \left( {\cal U}^{R \, k}_{2m} \right)^* \pm 
{\cal U}^{R \, k}_{1m}  \left( {\cal U}^{L\, k}_{2m} \right)^* \right) M_m \left(m_\mu \pm m_e \right) {\cal I}_{20}(r_{km})
\biggr\}  \nonumber \; ,
\end{eqnarray}
and
\begin{eqnarray}
\nonumber
\label{infEM1}
f_{M1,E1}(q^2)  \approx & -& \frac{m_\mu}{32 \pi^2} \sum_{k,m}
 \biggl\{  \frac{1}{M_m^2} \left( m_\mu \pm m_e \right) \left( {\cal U}^{L \, k}_{1m} \left( {\cal U}^{L \, k}_{2m} \right)^* \pm 
{\cal U}^{R \, k}_{1m}  \left( {\cal U}^{R\, k}_{2m} \right)^* \right) {\cal I}(r_{km}) \\ \nonumber
&& \hspace{1.5cm} + \frac{1}{M_m} \left( {\cal U}^{L \, k}_{1m} \left( {\cal U}^{R \, k}_{2m} \right)^* \pm 
{\cal U}^{R \, k}_{1m}  \left( {\cal U}^{L\, k}_{2m} \right)^* \right) {\cal J}(r_{km})
\biggr\} \\ \nonumber
& - & \frac{m_\mu q^2}{32 \pi^2} \biggl\{ \frac{1}{M_m^4} \left(m_\mu \pm m_e \right)  \left( {\cal U}^{L \, k}_{1m} \left( {\cal U}^{L \, k}_{2m} \right)^* \pm 
{\cal U}^{R \, k}_{1m}  \left( {\cal U}^{R\, k}_{2m} \right)^* \right) {\cal I}_{40}(r_{km}) \\ 
&& \hspace{1cm} + \frac{1}{M_m^3}  \left( {\cal U}^{L \, k}_{1m} \left( {\cal U}^{R \, k}_{2m} \right)^* \pm 
{\cal U}^{R \, k}_{1m}  \left( {\cal U}^{L\, k}_{2m} \right)^* \right) {\cal I}_{50}(r_{km})
\biggr\} \; .
\end{eqnarray}
Here $r_{km} = m^2_k/M^2_m$ and 
the expressions for the Feynman parameterization integrals  
${\cal I},\; {\cal J}$ and ${\cal I}_{i0} \, (i=1,2 ,\cdots,5)$ can be found in Appendix B.

From (\ref{convrate}) in Appendix B, 
the conversion rate (for $\gamma$ exchange) is given by
\begin{eqnarray}
\Gamma_{\rm conv} =
\frac{m_\mu^5}{4 \Lambda^4} \left( \biggl \vert C_{DR} D  
+ 4 \tilde C^{(p)}_{VR} V^{(p)} 
\biggr\vert^2 \, + \,  \biggl\vert C_{DL} D  + 4 \tilde C^{(p)}_{VL} V^{(p)} 
\biggr\vert^2  \right) \; ,
\label{convrategamma1}
\end{eqnarray}
where $C_{DR,DL}$ is given by (\ref{CsubDRDL}), 
and $\tilde C^{(p)}_{VR,VL}$ are given by (\ref{CsuppnsubVR}) 
and (\ref{CsuppnsubVL}) in Appendix A.
To obtain (\ref{convrategamma1}), we have used the following result valid for the neutron,
\begin{equation}
\tilde C^{(n)}_{V(L,R)}= \sum_{u, d, s} C^{(q)}_{V(L,R)} f^{(q)}_{Vn} = 0 \; .
\end{equation}
Using the above approximate form factors (\ref{infEM}) and (\ref{infEM1}) for small $q^2$,
we can derive
\begin{eqnarray}
\nonumber
C_{DR,DL}& \approx & \frac{e \Lambda^2}{32 \pi^2 m_\mu} \sum_{k,m} \biggl \{
\frac{{\cal I}(r_{km})}{M_m^2} \left(m_\mu {\cal U}^{R,L \, k}_{1m} \left( {\cal U}^{R,L \, k}_{2m} \right)^* + m_e
{\cal U}^{L,R \, k}_{1m}  \left( {\cal U}^{L,R\, k}_{2m} \right)^* \right) \nonumber \\
&& \hspace{60pt} + \frac{{\cal J}(r_{km})}{M_m} {\cal U}^{R,L \, k}_{1m}  \left( {\cal U}^{L,R\, k}_{2m} \right)^* \nonumber \\
&& \hspace{60pt} + \frac{q^2}{M_m^2} \bigg[
\frac{{\cal I}_{40}(r_{km})}{M_m^2} \left(m_\mu {\cal U}^{R,L \, k}_{1m} \left( {\cal U}^{R,L \, k}_{2m} \right)^* + m_e
{\cal U}^{L,R \, k}_{1m}  \left( {\cal U}^{L,R\, k}_{2m} \right)^* \right) 
\nonumber \\
&& \hspace{120pt} + \frac{{\cal I}_{50}(r_{km})}{M_m} {\cal U}^{R,L \, k}_{1m}  \left( {\cal U}^{L,R\, k}_{2m} \right)^* \bigg] 
\biggr \} \; ,
\end{eqnarray}
and summing over the contributions from light quarks, we have
\begin{eqnarray}
\nonumber
\tilde C^{(p)}_{VL,VR} \approx \frac{e^2 \Lambda^2}{16 \pi^2 M_m^4} \sum_{k,m}  &\biggl \{&
M_m^2 \left( {\cal I}(r_{km})+2 \;  {\cal I}_{30}(r_{km}) \right) {\cal U}^{R,L \, k}_{1m} \left( {\cal U}^{R,L \, k}_{2m} \right)^* \nonumber \\
&+ &m_\mu m_e {\cal I}_{10}(r_{km}) \; {\cal U}^{L,R \, k}_{1m} \left( {\cal U}^{L,R \, k}_{2m} \right)^* \\ \nonumber
&+& M_m {\cal I}_{20}(r_{km}) \left(m_\mu {\cal U}^{R,L \, k}_{1m} \left( {\cal U}^{L,R \, k}_{2m} \right)^* + m_e {\cal U}^{L,R \, k}_{1m} \left( {\cal U}^{R,L \, k}_{2m} \right)^* \right)
\biggr \} \; .
\end{eqnarray}
Dropping the $q^2$ terms in $C_{DR,DL}$ 
and keeping only those terms up to 
${\cal O}(1/M_m^2)$ in $\tilde C^{(p)}_{VL,VR}$,  
we obtain for the conversion rate 
\begin{eqnarray}
\Gamma_{\rm conv} (q^2 \to 0)& \approx &
\frac{m_\mu^5}{4} \frac{1}{(32 \pi^2)^2} \nonumber \\
&\times &\sum_{k,m} \biggl \{ 
 \biggl \vert \frac{16 \pi^2 D}{m_\mu} C_L^{km} +8 V^{(p)} e^2 \frac{{\cal I}(r_{km}) +2 \; {\cal I}_{30}(r_{km})}{M_m^2} {\cal U}^{L \, k}_{1m} \left( {\cal U}^{L \, k}_{2m} \right)^*  
\biggr\vert^2  \nonumber \\
&& + 
\biggl\vert \frac{16 \pi^2 D}{m_\mu} C_R^{km} +8 V^{(p)} e^2 \frac{{\cal I}(r_{km}) +2 \; {\cal I}_{30}(r_{km})}{M_m^2} {\cal U}^{R \, k}_{1m} \left( {\cal U}^{R \, k}_{2m} \right)^*
\biggr\vert^2 
\biggr \} \, ,
\label{convrategamma2}
\end{eqnarray}
where
\bea
\label{CLR}
C^{km}_{L,R} & = & \frac{e}{16 \pi^2} \biggl \{
\frac{{\cal I}(r_{km})}{M_m^2} \left(m_\mu {\cal U}^{R,L \, k}_{1m} \left( {\cal U}^{R,L \, k}_{2m} \right)^* + m_e{\cal U}^{L,R \, k}_{1m}  \left( {\cal U}^{L,R\, k}_{2m} \right)^* \right) \nonumber \\
&& \hspace{30pt} + \frac{{\cal J}(r_{km})}{M_m} {\cal U}^{R,L \, k}_{1m}  \left( {\cal U}^{L,R\, k}_{2m} \right)^*\biggl \} \; .
\eea
Recall that for the on-shell process $\mu \to e \gamma$, we have
\be
\Gamma_{\mu \to e \gamma} =\dfrac{1}{16 \pi} m_\mu^3 \sum_{k,m} \left(\lvert{C_L^{km}}\rvert^2 + \lvert{C_R^{km}}\rvert^2 \right) \; .
\ee
Thus, one obtains
\begin{eqnarray}
\label{convrategamma3}
\Gamma_{\rm conv} (q^2 \to 0) & \approx & \pi D^2 \Gamma_{\mu \to e \gamma}  
+\frac{m_\mu^5}{(64 \pi^2)^2} \sum_{k,m} \biggl \{ 
2DV^{(p)} \left(8 \pi e \right)^2 \frac{{\cal I}(r_{km}) +2  {\cal I}_{30}(r_{km})}{m_\mu M_m^2} \nonumber \\
&& \hspace{80pt} \times \left(C_L ^{km}{\cal U}^{L \, k}_{1m} \left( {\cal U}^{L \, k}_{2m} \right)^* + \left(C_L^{km}\right)^* \left({\cal U}^{L \, k}_{1m} \right)^* {\cal U}^{L \, k}_{2m} \right. \nonumber \\
&& \hspace{90pt} + \left.  C_R^{km} {\cal U}^{R \, k}_{1m} \left( {\cal U}^{R \, k}_{2m} \right)^* + \left(C_R^{km}\right)^*\left({\cal U}^{R \, k}_{1m} \right)^* {\cal U}^{R \, k}_{2m} \right) \\
&+& \left(8V^{(p)}e^2 \frac{{\cal I}(r_{km}) +2  {\cal I}_{30}(r_{km})}{M_m^2} \right)^2 \left(
\lvert {\cal U}^{L \, k}_{1m} \left( {\cal U}^{L \, k}_{2m} \right)^* \rvert^2 +
\lvert {\cal U}^{R \, k}_{1m} \left( {\cal U}^{R \, k}_{2m} \right)^* \rvert^2 \right)  
\biggr \} \; . \nonumber
\end{eqnarray}
Note that since $C^{km}_{L,R}$ is scaled by $1/M_m$, the second and the third terms in (\ref{convrategamma3}) are suppressed by 
$1/M_m$ and $1/ M_m^2$ respectively, as compared with the first term. If one drops these two suppressed terms further, one obtains
a simple relation
\begin{equation}
\Gamma_{\rm conv} (q^2 \to 0) \approx \pi D^2 \Gamma_{\mu \to e \gamma} \; .
\end{equation}
Thus,
\begin{equation}
B_{\mu N \to eN} = \frac{\Gamma_{\rm conv}}{\Gamma_{\rm capt}} \approx \pi D^2 \frac{\Gamma_\mu}{\Gamma_{\rm capt}} B_{\mu \to e \gamma} \; ,
\label{mu2esimple}
\end{equation}
where $\Gamma_\mu$ is the total decay width of the muon.


\section{Numerical analysis}\label{sec4}

In our analysis, we adapt the same assumptions for the parameter space 
as was done in \cite{Hung:2015hra}. We summarize them as follows.
\begin{itemize}
 \item For the mass parameters, we take the masses of the singlet scalars $\phi_{kS}$ to be
  \be
   m_0: m_1 : m_2 : m_3 = M_S : 2M_S : 3M_S : 4M_S\, ,
  \ee
where the common mass $M_S$ is set to be 10 MeV;
and for the mirror lepton masses, we set
   \be
   M_m = M_{\rm mirror} + \delta_m
   \ee
where $\delta_1 = 0,\; \delta_2 = 10 \gev,\; \delta_3 = 20 \gev$ and 
the common mass $M_{\rm mirror}$ is varied in the range of $100-800 \gev$. 
Our results are insensitive to these choices as long as $m_k/M_m \ll 1$.

 \item Note that the relations $g_{2S} = (g_{1S})^*$ and $g'_{2S} = (g'_{1S})^*$ hold 
 due to the hermiticity of the neutrino Dirac mass matrix. However, 
 all the Yukawa couplings $g_{0S},\; g_{1S},\; g_{2S},\; g'_{0S},\; g'_{1S},$ and $g'_{2S}$ are assumed to be real. 
 
 \item 
 Out of the four mixing matrices, only the one $U_{\rm PMNS}$ associated with the left-handed SM fermions 
 are known. Following \cite{Hung:2015hra}, we will consider two scenarios below:
	\begin{itemize}
	\item Scenario 1: $U_{\rm PMNS}^M = {U'}_{\rm PMNS} = {U'}_{\rm PMNS}^M = U_{\rm CW}^\dagger$
	\item Scenario 2: $U_{\rm PMNS}^M = {U'}_{\rm PMNS} = {U'}_{\rm PMNS}^M = U_{\rm PMNS}$
	\end{itemize}
where $U_{\rm CW}$ is given by (\ref{UCW}).
For the PMNS mixing matrix, we will use the best fit result in (\ref{UPMNSbestfits}).
In the two scenarios that we are studying, our results do not depend sensitively on the mass hierarchies.

\item
We will study the following two cases for the Yukawa couplings.
\begin{enumerate}
\item 
$g_{0S} = g'_{0S}$ and $g_{1S} = g'_{1S} = 10^{-2} g_{0S}$. Hence the contributions 
from the $A_4$ triplet is small.
\item 
$g_{0S} = g'_{0S} = g_{1S} = g'_{1S}$. Both $A_4$ singlet and triplet terms carry the same weight.
\end{enumerate}

\end{itemize}

\begin{figure}[hbtp!]
\centering
\includegraphics[width=1\linewidth]{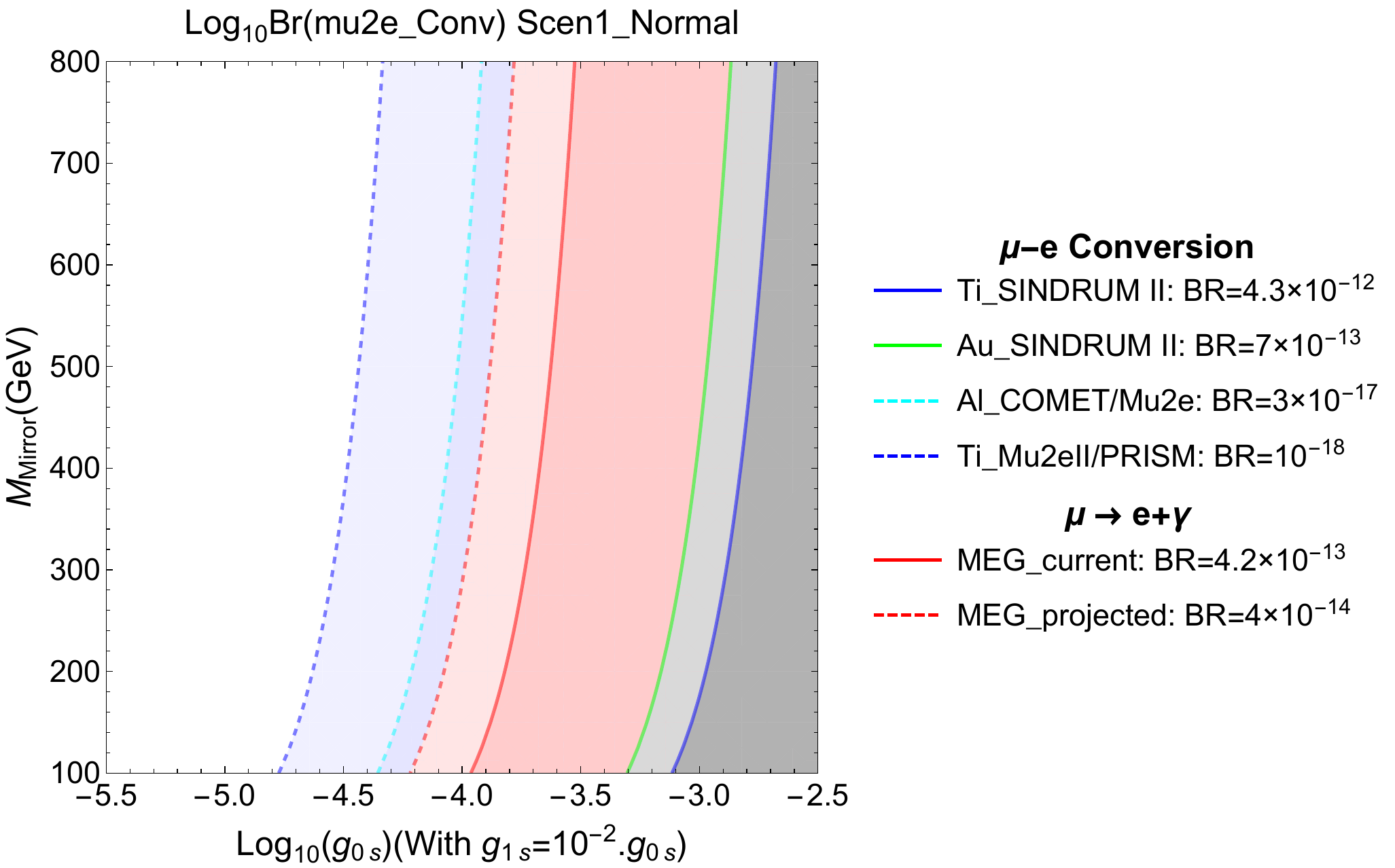} 
\caption{\small Contour plots of ${\rm Log}_{10} B(\mu-e \; {\rm conversion})$  
on the $(g_{0S},M_{\rm mirror})$ plane for normal mass hierarchy in Scenario 1 with $g_{0S}=g^\prime_{0S}$ and $g_{1S}=g^\prime_{1S}=10^{-2} g_{0S}$. The legend shows current experimental limits and projected sensitivities from COMET, Mu2e, SINDRUM II, PRISM and MEG.
For details of other input parameters, one can refer to the text in Sec.~\ref{sec4}.}
\label{fig2}
\end{figure}

\begin{figure}[hbtp!]
\centering
\includegraphics[width=1\linewidth]{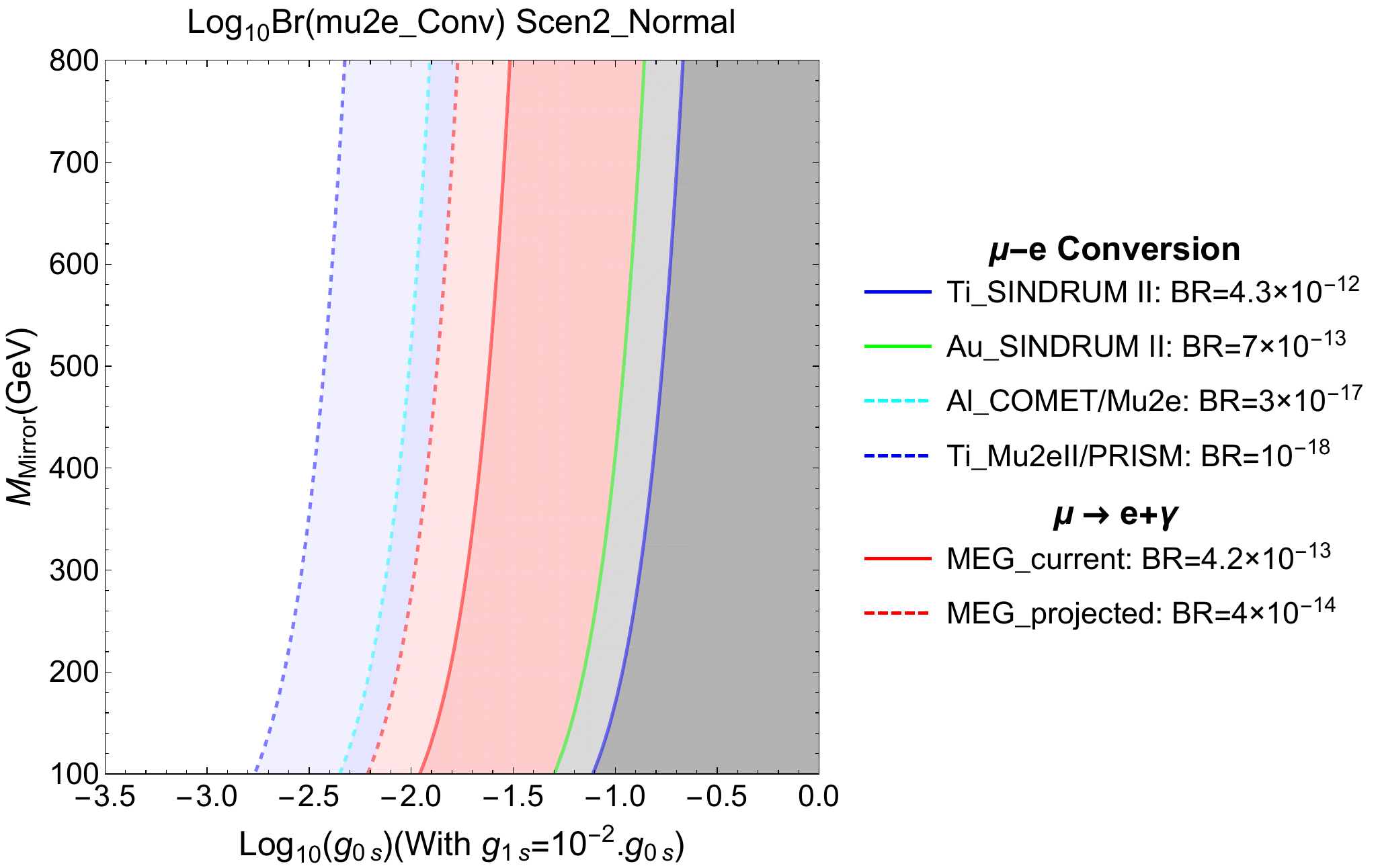} 
\caption{\small Contour plots of ${\rm Log}_{10} B(\mu-e \; {\rm conversion})$  
on the $(g_{0S},M_{\rm mirror})$ plane for normal mass hierarchy in Scenarios 2 with $g_{0S}=g^\prime_{0S}$ and $g_{1S}=g^\prime_{1S}=10^{-2} g_{0S}$.}
\label{fig3}
\end{figure}
\begin{figure}[hbtp!]
\centering
\includegraphics[width=1\linewidth]{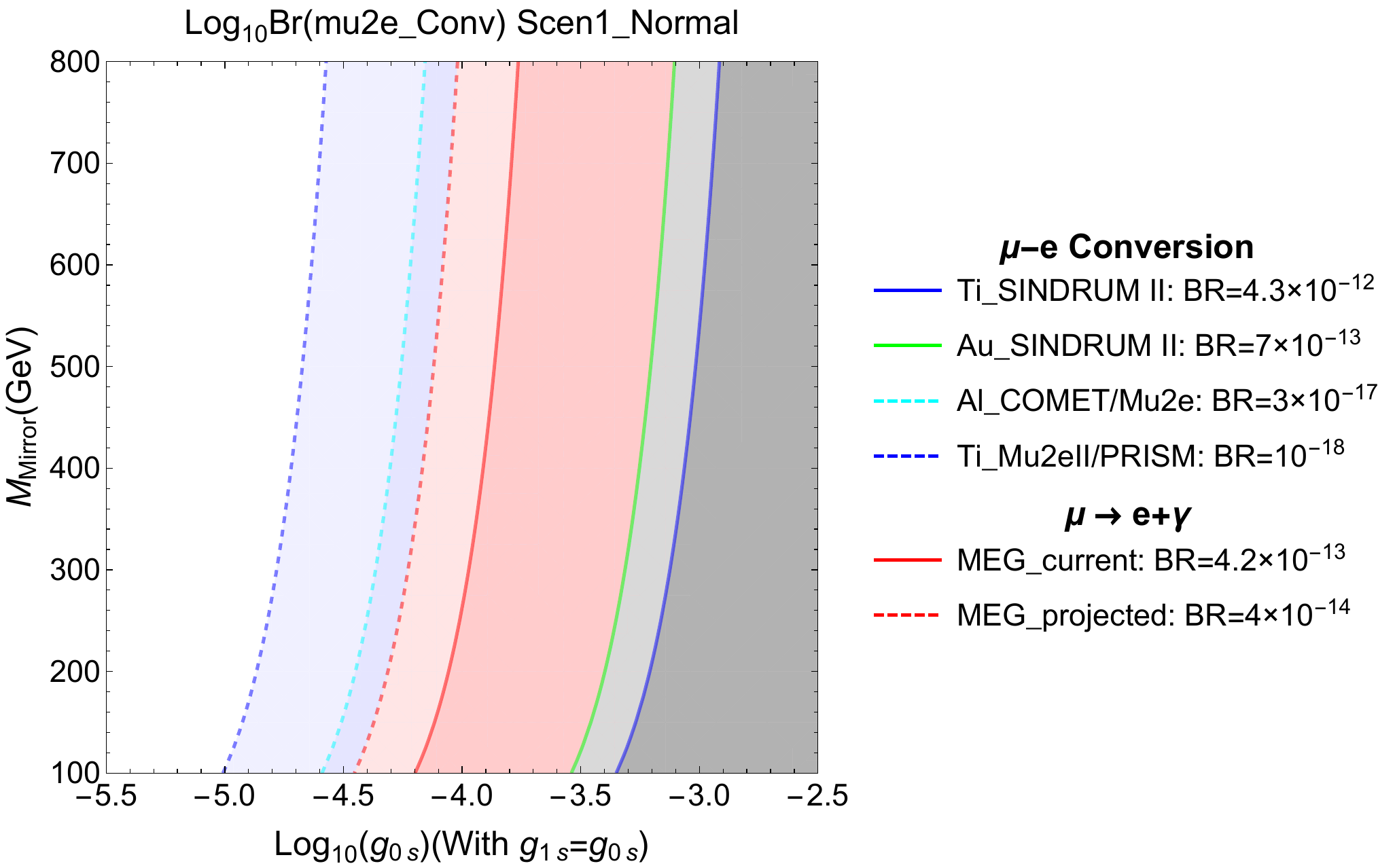} 
\caption{\small Contour plots of ${\rm Log}_{10} B(\mu-e \; {\rm conversion})$  
on the $(g_{0S},M_{\rm mirror})$ plane for normal mass hierarchy in Scenarios 1 with $g_{0S}=g^\prime_{0S}=g_{1S}=g^\prime_{1S}$}
\label{fig4}
\end{figure}
\begin{figure}[hbtp!]
\centering
\includegraphics[width=1\linewidth]{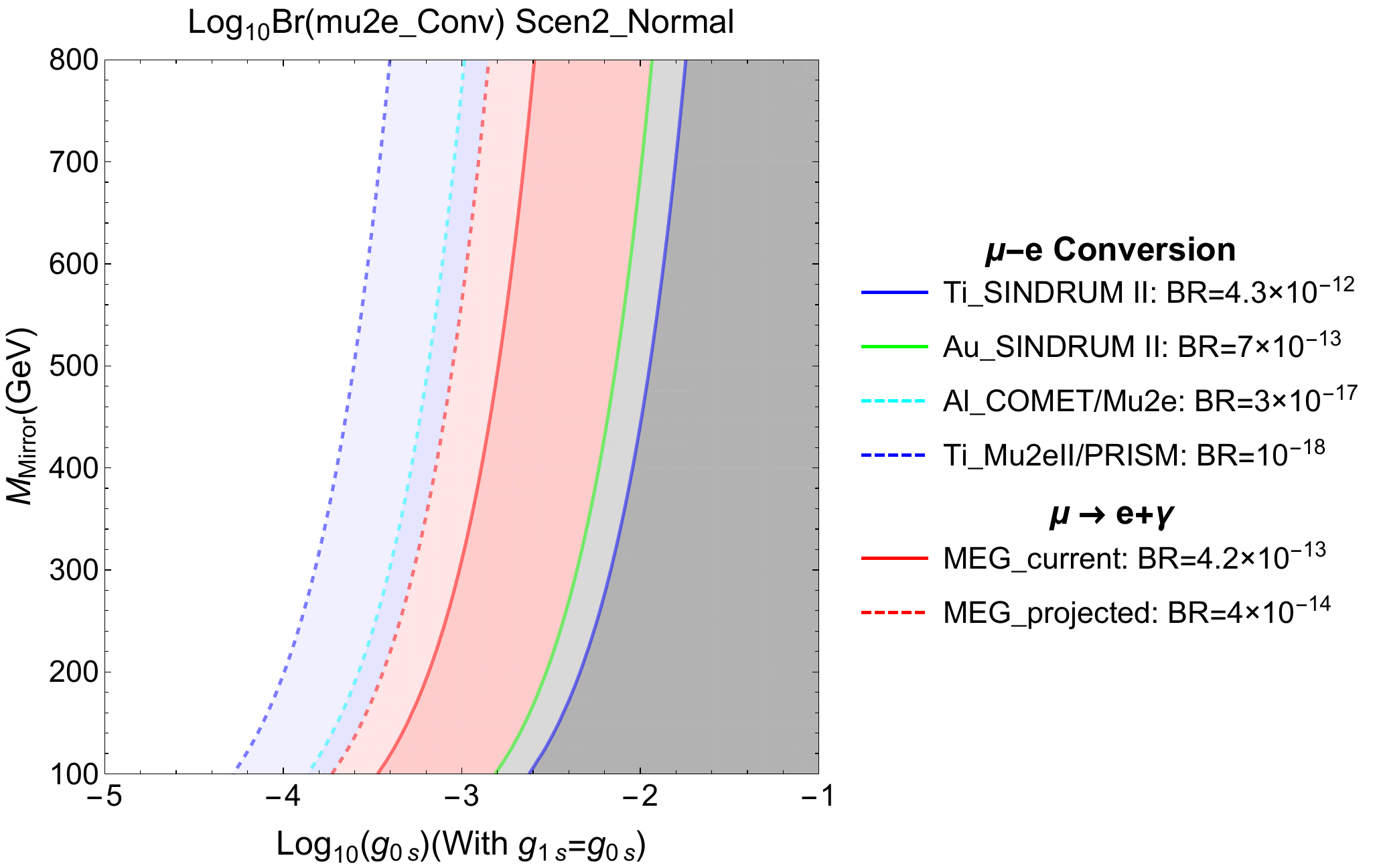} 
\caption{\small Contour plots of ${\rm Log}_{10} B(\mu-e \; {\rm conversion})$  
on the $(g_{0S},M_{\rm mirror})$ plane for normal mass hierarchy in Scenarios 2 with $g_{0S}=g^\prime_{0S}=g_{1S}=g^\prime_{1S}$}
\label{fig5}
\end{figure}
In Figs.~\ref{fig2}, \ref{fig3}, \ref{fig4} and \ref{fig5} we plot the contour of $\log_{10}B(\mu-e \; {\rm conversion})$ with $\gamma$ dominance in the $(\log_{10}(g_{0S}), M_{\rm mirror})$ plane for Scenarios 1 and 2 with the normal neutrino mass hierarchy for the 2 cases of couplings aforementioned respectively.
The blue and green solid lines correspond to the current limits from SINDRUM II experiments for $\mu - e$ conversion to titanium~(\ref{sindrum2Ti}) and gold~(\ref{sindrum2Au}) respectively.
The red solid and dashed lines correspond to the current limit~(\ref{megcurrent}) and projected sensitivity~(\ref{megprojected}) for $\mu \to e \gamma$ from MEG experiment.
The cyan and blue dashed  lines correspond to the projected sensitivities for $\mu - e$ conversion to aluminum and titanium from COMET, Mu2e~(\ref{COMETmu2eprojected}) and Mu2e II, PRISM ~(\ref{mu2eIIPRISMprojected}) experiments respectively.

Several comments are in order here.
\begin{itemize}
\item
We have studied in some details the effects of different settings of couplings on our results.
Generally, we observe that as one varies the $A_4$ triplet coupling $g_{1S}$ from $10^{-2} g_{0S}$ to $g_{0S}$ (from Figs.~\ref{fig2} to~\ref{fig5}) the contour plots for $\log_{10}B(\mu-e \, {\rm conversion})$ are shifted to the left. 
The $A_4$ triplet is playing a significant role in putting constraints on the parameter space for the CLFV processes, 
such as $\mu \to e \gamma$ and $\mu-e$ conversion in the model.
\item
For the sensitivity of the two scenarios, we find that 
 \begin{itemize}
 \item 
 Generally, Scenario 2 is less stringent constraint than Scenario 1.
 \item
 In particular, when the $A_4$ singlet couplings are dominated, Scenario 2 is less stringent than Scenario 1 by at least two order of magnitude ($10^{-3}$ vs. $10^{-1}$), regarding the constraint on the couplings (as shown in Figs.~\ref{fig2} and \ref{fig3}). This is due to the fact that in Scenario 2, the three unknown unitary mixing matrices are now departure from $U_{\rm PMNS}$ which allows for larger effects since the amplitudes involve products of both the couplings and the elements of mixing matrices.
 \item
 However, as one turns on the contribution from the $A_4$ triplet in Fig.~(\ref{fig4}) and Fig.~(\ref{fig5}), the discrepancy between two scenarios 1 and 2 shrink ($10^{-2}$ vs. $10^{-3.2}$). It implies that Scenario 2 is more sensitive to the change in the structure of $A_4$ couplings.
 \end{itemize}
\item
Finally, regarding the incorporation of the current limit on $B(\mu \to e \gamma)$ from MEG experiment and its projected sensitivity into the contour plots of $\log_{10}B(\mu-e \, {\rm conversion})$, one can obtain the following statements by looking at  Figs.~(\ref{fig2})-(\ref{fig5}):
\begin{itemize}
\item 
The plots illustrate nicely the close relation between the two CLFV processes 
$\mu \to e \gamma$ and $\mu-e$ conversion in nuclei
using the simple formula (\ref{mu2esimple}) we derived in Sec.~\ref{sec3}.
\item 
In the same parameter space, $\mu \to e \gamma$ shows a tighter constraint than $\mu-e$ conversion by the fact that it excludes almost half of the searched region for the branching ratio of $\mu-e$ conversion. Therefore, our work helps narrow down future searches for $\mu-e$ conversion at Fermilab/Mu2e, J-PARC/COMET and PRISM.
\item
With the current upper bounds from various experiments, the radiative decay
$\mu \to e \gamma$ is providing more stringent constraints on the 
couplings than the $\mu-e$ conversion ($10^{-4}$ vs. $10^{-3}$, about one order of magnitude better). 
However, for the future projected sensitivities at Mu2e and COMET, $\mu-e$ conversion is slightly more stringent, about half an order of magnitude stronger constraints on the couplings. For PRISM, it can be about an order of magnitude more stronger.
\end{itemize}
\end{itemize}

\section{Summary}
\label{sec5}

Mirror fermion model with electroweak scale non-sterile right-handed neutrinos is an interesting
extension of the SM. Aside from its aesthetically appealing to restoring parity symmetry 
at higher energy scale,
it can have immediate impacts for experiments in both complementary frontiers of high energy and high intensity searching for new physics of lepton flavor violation.

In this study, we discussed $\mu - e$ conversion in nuclei and radiative decay $\mu \to e \gamma$
in an extended mirror fermion model with a $A_4$ horizontal symmetry in the lepton sector.
Currently the most stringent constraint on the parameter space of the model is provided by the 
most recent limit on the radiative decay $\mu \to e \gamma$ 
from MEG. In the future, Mu2e and COMET experiments can provide more stringent constraints on the 
model from $\mu - e$ conversion in aluminum. 
The sensitivity of the new Yukawa couplings can be probed is of order $10^{-5}$, about 
one order of magnitude improvement compared with current status from MEG. 
Small Yukawa couplings such as $10^{-5}$ can give rise to distinct signatures in the search of mirror charged leptons and Majorana right-handed neutrinos at the LHC (or planned colliders) in the form of displaced decay vertices with decay lengths larger than 1 mm or so~\cite{Chakdar:2016adj}. Although unrelated to the present analysis, a similar remark can be made for the search for mirror quarks~\cite{Chakdar:2015sra}.

It would be interesting to extend this study to the electric dipole moment of the electron and neutron,
which requires the new Yukawa couplings to take on complex values instead of real ones as assumed in the present analysis.
This work is in progress and will be presented elsewhere~\cite{edmpaper}.


\vfill

\section*{Appendix A - Effective Lagrangian For $\mu - e$ Conversion}
\label{appendixa}

Effective Lagrangian is a powerful technique to analyze low energy processes 
like $\mu \to e$ conversion in nuclei since the momentum transfer $q^2$ is typically 
of the order ${\cal O}(m_\mu^2) \ll m^2_N$ for nucleus $N$.
The most general CLFV effective Lagrangian which contributes to the $\mu - e$ conversion
in nuclei has been studied by various groups~\cite{Kuno:1999jp,Kitano:2002mt,Cirigliano:2009bz}. 
At the scale $\Lambda$ where the heavy particles (including particles beyond the SM as well as 
the heavy top, bottom and charm quarks) being integrated out, the relevant terms for the model we are studying are
\begin{eqnarray}
{\cal L}_{\rm eff} = & - &\frac{1}{\Lambda^2} \biggl[ 
\left( C_{DR} m_\mu \overline e \sigma^{\alpha\beta} P_L \mu 
+ C_{DL} m_\mu \overline e \sigma^{\alpha\beta} P_R \mu \right) F_{\alpha\beta} \biggr. \nonumber \\
&+& \sum_{q=u,d,s}
\left( C^{(q)}_{VR} \overline e \gamma^\alpha P_R \mu + C^{(q)}_{VL} \overline e \gamma^\alpha P_L \mu \right)
\overline q \gamma_\alpha q + {\rm H.c.} \biggr] \; .
\label{Leff2}
\end{eqnarray}
Here $m_\mu$ is the muon mass;
$P_{L,R}=(1\mp\gamma_5)/2$, $\sigma_{\mu\nu}=i \left[ \gamma_\mu , \gamma_\nu \right]/2$;
$F_{\alpha\beta}$ is the electromagnetic field strength;
finally, $C_{D(L,R)}$ and $C^{(q)}_{V(L,R)}$ are dimensionless coupling constants depending on specific LFV model. 
In the specific mirror model calculation, 
we will be focusing on the photon and $Z$ boson exchange diagrams 
which contribute only to the magnetic and electric dipole moment operators
as well as the vector and axial vector lepton bilinears. 

To determine the conversion rate, the above effective Lagrangian (\ref{Leff2}) is needed to scale down to the 
nuclear scale where the hadronic matrix elements
$\langle N \vert \overline q \gamma_\mu q\vert N \rangle$,
$\langle N \vert F^{\alpha\beta}F_{\alpha\beta} \vert N \rangle$ are evaluated.
In addition, the muon and electron wave functions may be significantly deviated from plane wave due to 
distortion by the coulomb potential of the nuclei. For high $Z$ nuclei, relativistic corrections to their wave functions 
are important as well. 
The formula for the conversion rate is given by \cite{Kitano:2002mt,Cirigliano:2009bz}
\begin{eqnarray}
\Gamma_{\rm conv} &=&
\frac{m_\mu^5}{4 \Lambda^4} \left( \biggl \vert C_{DR} D  
+ 4 \tilde C^{(p)}_{VR} V^{(p)} +  4 \tilde C^{(n)}_{VR} V^{(n)} 
\biggr\vert^2  +  \biggl\vert C_{DL} D  
+ 4 \tilde C^{(p)}_{VL} V^{(p)} +  4 \tilde C^{(n)}_{VL} V^{(n)} 
\biggr\vert^2  \right) \; .  \nonumber \\
\label{convrate}
\end{eqnarray}
In (\ref{convrate}) the coupling constants $\tilde C^{(p,n)}_{V(R,L)}$  are defined as \cite{Cirigliano:2009bz}
\begin{eqnarray}
\label{CsuppnsubVR}
\tilde C^{(p)}_{VR} &=& \sum_{q=u,d,s} C^{(q)}_{VR} f^{(q)}_{Vp}  \; , \\ 
\tilde C^{(n)}_{VR} &=& \sum_{q=u,d,s} C^{(q)}_{VR} f^{(q)}_{Vn}  \; , \\ 
\label{CsuppnsubVL}
\tilde C^{(p)}_{VL} &=& \sum_{q=u,d,s} C^{(q)}_{VL} f^{(q)}_{Vp}  \; , \\ 
\tilde C^{(n)}_{VL} &=& \sum_{q=u,d,s} C^{(q)}_{VL} f^{(q)}_{Vn}  \; ,
\end{eqnarray}
where $f^{(q)}_{Vp}$ and $f^{(q)}_{Vn}$ are the known nucleon vector form factors
\begin{equation}
\begin{array}{lll}
f^{(u)}_{Vp} = 2, & f^{(d)}_{Vp} = 1, & f^{(s)}_{Vp} = 0 \; , \\
f^{(u)}_{Vn} = 1, & f^{(d)}_{Vn} = 2, & f^{(s)}_{Vn} = 0 \; . 
\end{array}
\end{equation}

\begin{table}[t!]
\caption{Values of the dimensionless overlap integrals for aluminum, titanium and gold, 
evaluated under the assumption that the proton and neutron distributions within each nuclei 
are the same~\cite{Kitano:2002mt}.}
\begin{tabular}{lccc}
\hline
Nucleus & $D$ &  $V^{(p)}$ &  $V^{(n)}$ \\
\hline\hline
$^{27}_{13}$Al & 0.0362 & 0.0161 & 0.0173 \\
$^{48}_{22}$Ti & 0.0864 & 0.0396 & 0.0468  \\
$^{197}_{79}$Au & 0.189 & 0.0974 & 0.146 \\
\hline
\end{tabular}
\label{overlapintegrals}
\end{table}
The dimensionless quantities
$D$ and $V^{(p,n)}$ in (\ref{convrate}) are the overlap integrals of 
the relativistic wave functions of muon and electron in the electric field of the nucleus 
weighted by appropriate combinations of proton and neutron densities \cite{Kitano:2002mt}. 
Their values for the four nuclei aluminum, titanium, gold and lead 
are listed in Table~\ref{overlapintegrals} for reference.

The $\mu - e$ conversion branching ratio is defined as
\begin{equation}
B_{\mu N \to e N}(Z,A) \equiv \frac{\Gamma_{\rm conv}}{\Gamma_{\rm capt}} \; ,
\end{equation}
where $\Gamma_{\rm conv}$ is given by (\ref{convrate}) and  $\Gamma_{\rm capt}$ is the standard model 
muon capture rate. The SM capture rates for aluminum, titanium and gold have been 
determined experimentally \cite{muoncapturerate}
and they are listed in Table~\ref{capturerate} for convenience.

\begin{table}[t!]
\caption{Standard model values of the capture rates for aluminum, titanium and gold
in unit of $10^6 \, {\rm s}^{-1}$ taken from Ref.~\cite{muoncapturerate}. }
\begin{tabular}{lc}
\hline
Nucleus & $\Gamma_{\rm capt}$ ($10^{6}$ s$^{-1}$)  \\
\hline\hline
$^{27}_{13}$Al &  0.7054  \\
$^{48}_{22}$Ti &  2.59 \\
$^{197}_{79}$Au & 13.07 \\
\hline
\end{tabular}
\label{capturerate}
\end{table}

\vfill

\section*{Appendix B - 
Formulas for ${\cal I}, {\cal J}, {\cal I}_{i0} (i=1,\cdots,5)$}
\label{appendixb}

In the limit of zero momentum transfer, the Feynman parameterization integrals 
in the various form factors can be carried out analytically.
We collect their results here.
\bea
\label{I}
{\cal I}(r) & = & \frac{1}{12\, (1 - r)^4} \left[ - 6 r^2 \log r + r ( 2 r^2 + 3 r - 6 ) + 1 \right] \; , \\
\label{J}
{\cal J}(r) & = & \frac{1}{2 \,(1 - r)^3} \left[ - 2 r^2 \log r + r ( 3 r - 4) + 1 \right] \;  , \\
\label{I10}
{\cal I}_{10}(r) &=& \frac{1}{72\,(1-r)^6} \left[-12r^2(3+2r) \log r  +(r-1) \left(3r^3+47r^2+11r-1 \right) \right] \;  , \\
\label{I20}
{\cal I}_{20}(r) &=& \frac{1}{36\,(1-r)^5} \left[-6r^2(3+r) \log r +(r-1) \left(17r^2+8r-1 \right) \right] \;  , \\
\label{I30}
{\cal I}_{30}(r) &=& \frac{1}{36\,(1-r)^4} \left[6r^3 \log r + \left(-11r^2+18r-9 \right)r+2 \right] \;  , \\
\label{I40}
{\cal I}_{40}(r) &=&  \frac{1}{144\, (1-r)^6} \left[12r^3(r+4) \log r - \left(r^2-1\right) \left(37r^2-8r+1 \right) \right] \;  , \\
\label{I50}
{\cal I}_{50}(r) &=&  \frac{1}{18 \,(1-r)^5} \left[12r^3 \log r -  \left(3r^4+10r^3-18r^2+6r \right) +1 \right] \; .
\eea
Here $r$ denotes the mass ratio $m^2/M^2$, where $m$ and $M$ are the masses of the scalar singlet and mirror lepton respectively.


\vfill

\section*{Acknowledgments}
TCY would like to thank T. N. Pham for useful discussions and 
the hospitality he received at Centre de Physique Th${\acute {\rm e}}$orique de 
l'Ecole Polytechnique  where progress of this project was made.
We would also like to thank Craig Group for useful comments on the manuscript. 
This work was supported in part by the Ministry of Science and Technology (MoST) of Taiwan under
grant number 104-2112-M-001-001-MY3.


\end{document}